\documentclass[10pt]{article}

\usepackage[dvips]{graphicx}

\usepackage[authoryear, round, sort]{natbib}

\usepackage{amssymb}
\usepackage{amsfonts}
\usepackage{amsmath}
\usepackage{color}
\usepackage{enumerate}
\usepackage{times}
\usepackage[T1]{fontenc}
\usepackage{amsmath}
\usepackage{graphics}
\usepackage{epsfig} 
\usepackage[utf8]{inputenc}

\usepackage{lscape}

\usepackage{longtable}

  \def\wVar{\mathop{\rm wVar}}
    \def\wCov{\mathop{\rm wCov}}
  \def\CEI{\mathop{\rm CEI}}

\usepackage[labelfont=bf,labelsep=period,justification=raggedright]{caption}

\makeatletter
\renewcommand{\@biblabel}[1]{\quad#1.}
\makeatother

\date{}
\begin{document}

\begin{flushleft}
{\Large
\textbf{Citation impacts revisited: 
how novel impact measures reflect interdisciplinarity and structural change at the local and global level}
}
\\
Michel Zitt$^{1}$$\ast$, 
Jean-Philippe Cointet$^{2}$$^{3}$
\\
\bf{1 }{\sc INRA-Lereco, Nantes, France}
\\
\bf{2} {\sc INRA-SenS, Marne-la-Vallée, France}
\\
\bf{3} {\sc ISC-PIF, Paris, France}
\\
$\ast$ E-mail:  Michel.Zitt@nantes.inra.fr
\end{flushleft}


.

\vspace{-.5cm}
\section*{Abstract}
Citation networks have fed numerous works in scientific evaluation, science mapping (and more recently large-scale network studies) for decades.  The variety of citation behavior across scientific fields is both a research topic in sociology of science, and a problem in scientific evaluation. Normalization, tantamount to a particular weighting of links in the citation network, is necessary for allowing across-field comparisons of citation scores and interdisciplinary studies. In addition to classical normalization which drastically reduces all variability factors altogether, two tracks of research have emerged in the recent years. One is the revival of iterative "influence measures". The second is the "citing-side" normalization, whose only purpose is to control for the main factor of variability, the inequality in citing propensity, letting other aspects play: knowledge export/imports and growth. When all variables are defined at the same field-level, two propositions are established: (a) the gross impact measure identifies with the product of relative growth rate, gross balance of citation exchanges, and relative number of references (b) the normalized impact identifies with the product of relative growth rate and normalized balance. At the science level, the variance of growth rate over domains is a proxy for change in the system, and the variance of balance a measure of inter-disciplinary dependences. This opens a new perspective, where the resulting variance of normalized impact, and a related measure, the sum of these variances proposed as a Change-Exchange Indicator, summarize important aspects of science structure and dynamism. Results based on a decade's data are discussed. The behavior of normalized impact according to scale changes is also briefly discussed. A shift towards a network-based definition of domains, more in the nomenclature-free spirit of citing-side normalization than database classification schemes, appears promising, albeit with technical challenges. An appealing issue is the connection with macro-level life-cycles of domains, and the dynamics of citation network.

\section*{Introduction }
The use of citation measures in science \citep{Garfield:1955tm,Garfield:2006gm} is a controversial issue in research evaluation, as shown in the recurrent debates on impact factors\footnote{In recent literature, see the dedicated issue of \textit{Scientometrics}, 92(2), (2012)}. Citations also shape a large-scale network \citep{deSollaPrice:1965vs,redn:cita} which, along with collaboration, linguistic and web-communication networks, is a powerful tool for mapping science and understanding knowledge exchanges and self-organization of communities. A lasting issue is the variability of citation practices across fields, which prevents any sensible comparison between gross citation figures or h-indexes, say in mathematics vs. cell biology. A traditional way to deal with this variability is the normalization of citation figures based on fields baseline figures \citep{murugesan2007variation,schubert1986relative,sen1992normalised,Czapski:1997jh,vinkler2002subfield,radicchi2008universality}. This "ex post" or "cited-side" statistical normalization is typically nomenclature-dependent, assuming an explicit delineation of scientific domains, usually from databases classification schemes. Forcing equality of cited domains, it sacrifices the consistency of the network and jeopardises multidisciplinary analysis. An alternative is the citing-side normalization ("ex ante", "source-level", "fractional citation";  \citep{zitt2008modifying,moed2010measuring,Glanzel:2011df,Leydesdorff:2010wb,waltman2012inconsistency}. The citing-side perspective \citep{zitt2005relativity} is at the confluence of Garfield's insights on citation density \citep{Garfield:1979uz} and fractional weighting to reduce biases in cocitation mapping \citep{small1985clustering}.  It corrects for the unequal propensity to cite amongst domains: in doing so, it keeps the best part of normalization – by removing undesirable sources of across-fields variability – while keeping the coherence of the citation network.

Especially, the partial normalization brought by the citing-side process, gives interpretable figures of domain-level average impact, which is true neither for usual "cited-side" normalized figures, forced to equality, nor for gross citation figures, blurred in magnitude by the effects of differential propensity to cite amongst fields. Focusing here on the analysis at the aggregate levels, we argue that citing-side approach opens new perspectives on interpretation of citation impacts at the domain level, and on structure and change of science insofar as it can be depicted by citation networks. We shall first establish two basic propositions on the decomposition of gross impacts and citing-side normalized impacts at the domain level. For the latter, we propose to summarize into a "Change-Exchange Index" the variances over domains of its two factors at the domain level, growth rate and dependence. It may seem strange at first  to come across a time-dependent variable such as growth, but the diachronic nature of citations implicitly carries information on change.

The $\CEI$ identifies with the variance of normalized impact when the two factors are independent, making the covariance term zero. We shall examine, on a decade's data on citation flows across science and a fixed nomenclature of domains, the empirical value of the variance and covariance terms calling for interpretations in terms of dynamics of the system, and discuss the challenges of shifting the nomenclature-based analysis to a bibliometric classification into topic/domains, more in line with the nomenclature-free spirit of citing-side approach.  

In section 1, analytical, we shall state two propositions at the domain level: one on gross impacts,one on citing-side normalized impacts. Then considering the science level, we will define the  and its relation with the normalized impact. Section II summarizes first results from an on-going empirical analysis on a decade of the Web of Science. The discussion section discusses several aspects: the shift from a database classification scheme (nomenclature) framework to a bibliometric classification of science; the relationship with various aspects of dynamics of science. 

\section{Analytical bases}
If all variables are calculated at the same level of classification (whatever the level: for example the subject category) we get two basic propositions. 
\paragraph{proposition 1}- domain level, gross impact (not normalized) 
\medskip

The impact  $I(A)$ of a domain $A$ is defined as the average number of incoming citations per articles susceptible to get cited in $A$. If $\Phi_\leftarrowtail(A)$ denotes the aggregated number of references citing domain $A$  then : $I(A)=\frac{\Phi_\leftarrowtail(A)}{|A|}$.
The growth rate $\rho(A)$ of a domain $A$ is simply defined here by the ratio of publication volumes between the cited and the citing periods, volumes reduced to average volume over each period. We then introduce the balance which compares the total inflow of citation with total outflow emitted by $A$ ($B(A)=\frac{\Phi_\leftarrowtail(A)}{\Phi_\rightarrowtail(A)}$). Finally we denote $\kappa(A)$ the average number of references in citing articles in $A$.
It is then straightforward to deduce the following equation (seen Appendix for further details):
\begin{equation}
I(A)=\rho(A)B(A)\kappa(A)
\end{equation}

From this equation, it is useful to introduce the notation $.\hat{}$ ~ transforming any domain level index into its relative version normalized with its science level counterpart. Given any domain-level measure $m(A)$ one can compute $\hat{m}(A)=\frac{m(A)}{m(S)}$. Thus the relative impact $\hat{I}(A)$ is obtained by dividing the gross impact by the impact computed at the whole science  level ($I(S)$). We will also denote  the relative growth rate $\hat{\rho}(A)$ (\textit{i.e.} growth rate normalized by the growth rate at the global science level) and $\hat{\kappa}(A)$ the relative number of references in citing articles in $A$.

\begin{equation}
\hat{I}(A)=\hat{\rho}(A)B(A)\hat{\kappa}(A)
\end{equation}
Proof is given in Appendix.

\paragraph{proposition 2} - domain level: citing-side normalized impact
\medskip

In order to neutralize the main source of variability, a normalization based on the relative number of active references (the "citing propensity") is introduced. It is implemented by weighting the links of the original directed and unweighted citation network, with options fixing the granularity of the baseline. In a simple device, cited-side normalization weighs links  proportionally to average in-links by node within the citable set in a given domain's delineation while citing-side normalization weighs links proportionally to average out-links by node within the citing set in the domain. Those domains can be defined by some neighbourhood of the citing article: journal, cluster, or librarians/database categories. Here, for establishing basic propositions, we shall rely on subject categories as defined by Web of Science (Thomson Reuters). With such a weighting of the citation links it naturally appears that $)\kappa(A)=1$. Neutralizing citing propensity variability then defines a new measure of impact which can be decomposed as: 

\begin{equation}
\hat{I}_g(A)=\hat{\rho}(A)B_g(A)
\end{equation}
These propositions generalize previous results on the journal impact factor \citep{zitt2011behind}. 

\paragraph{proposition 3} - science level: the deviation of citation impacts. 
\medskip

If the domain-level normalized impact is the product of two relative measures linked to interdisciplinary structure (asymmetry of exchange) and local dynamism (relative growth), what can we learn at the science level? All measures being relative, the signs of change are expected in the deviation indexes.  We shall limit ourselves to the variances (on the log-transformed variables), in spite of imperfections, but concentration indexes such as the Gini mean difference \cite{yitzhaki2003gini} with larger scope of application could be envisioned. 

For a particular category $A$ at a given level of breakdown $\hat{I}_g(A)=\hat{\rho}(A)B_g(A)$. With logarithmic transformation of variables, suggested by the distribution of impacts at the domain level:
$LI(A)=LG(A)+LB(A)$ where $LI$, $LG$, $LB$  designate respective logs of normalized impact, growth rate and normalized balance. Over all domains:

\begin{equation}
\wVar(LI)=\wVar(LG)+\wVar(LB)+2\wCov(LG,LB)
\end{equation}
where $\wVar$ stands for variance weighted by the volume of publications of domains, expressed in number of publications. For comparison sake, the unweighted variance has also been used.

In Equation 4 the variance terms have a simple interpretation. $\wVar(LB)$ over domains is a proxy of global interdisciplinary dependences in the system, and $\wVar(LG)$  is a proxy for the intensity of "creative destruction" through differentiation of growth rates over domains. A scientific system where domains do not exchange and are in steady state will associate zero variance and covariance terms, giving a zero variance of impacts. At the opposite end, a scientific system combining a high proportion of growing and declining domains and a strongly asymmetrical balance of flows across fields (exporters and importers) will show a high level of variance terms, but the final value of $\wVar(LI)$  will also depend on the covariance term.

The relationship between growth and balance partly depend on the superposition of domains at various stages of their life-cycle, while the potential value of balance for individual domains, typically reached at maturity stages, can show great dispersion linked to the position of the domain in the cognitive chain. The variance of balance (compared to growth's) may play a dominant role in the shaping of impact dispersion . Domains in emergence both grow rapidly and are quite dependent on imports of knowledge/information from their parent fields. Hence they are likely to enhance the variance of growth, and to yield negative covariance

In order to summarize asymmetry and growth effects, we propose then to consider only the sum of variance terms, the "structural-change and exchange-asymmetry index", abridged into Change-Exchange Index, $\CEI$ :

\begin{equation}
\CEI=\wVar(LG)+\wVar(LB)
\end{equation}
This index is closely related to the variance of impacts with $\CEI = \wVar(LI)-2\wCov(LG,LB)$.

$\CEI$ is trivially equal to the variance of impacts if growth rate and balance are independent.
scale issues
If the level of calculation of impact and the level of normalization (at which balances and growth rate are computed) are different, factors of scale come into play. Let us for example compare the normalized impact of sub-disciplines obtained (a) by normalization on the same-level, sub-discipline (b) by normalization at inferior level, the subject category. The growth factor for (a) is the weighted mean of growth factor for the corresponding categories.For the balance factor (b), a correcting coefficient depending on the structure of exchanges is needed, since the global balance of say a discipline is obviously not the average of the balances of the component categories. As far as gross impacts are concerned: the impact, the growth factor and the relative length of bibliographies are stable in aggregations with appropriate weighting by the volume of publications, whereas gross balances are not.  Scale irregularities in standard (cited-side) normalization had also already been stressed by \cite{zitt2005relativity}.

In such configurations where the level of definition of impacts and of other variables are not homogeneous (which is the case in many practical uses of normalization), the relations above should be altered by a correcting factor for the balance.

\section{A first experiment within a fixed nomenclature}
Data are based on OST aggregate figures at the category level, based on primary data and subject categories from the Web of Science (Thomson Reuters). 

The citation framework is based on "cited years", on the period 1999-2010, giving an exploitable span 1999-2006 and with caution through 2008 (with reduced but acceptable citation window). In the database (OST-WoS), there are overlaps in assignment of journals and then papers to categories (WoS subject categories) at the lower level, handled by fractional counting. The nomenclature at the sub-discipline and the discipline level is derived from OST scheme, modified for simplicity sake, in order to get an embedded scheme:

\begin{center}specialty (subject category) $\subset$  sub-discipline $\subset$  discipline
\end{center}

The nomenclature covers all sciences including social sciences and humanities. Specialties with very low citations activity, most of them belonging to humanities where the interpretation of citations is problematic, were discarded\footnote{The resulting selection is given in appendix II}. Let us summarize the main results:
\paragraph{Gross impact}: As expected the gross impact heavily depends on the propensity to cite. The variance of the impact is essentially shaped by the variance of this factor, which jeopardizes any interpretation of its variance in terms of balance, the variance of which is by an order of magnitude lower, and growth, still behind.

\paragraph{Normalized impact}: As soon as citing propensity is corrected, a new avenue is open to interpretation of citation impacts, in terms of dynamism and asymmetrical interdisciplinarity. Fig. 1 shows the time series of variance (weighted variance) of normalized impact, of its factors growth and balance, of the covariance term, and the series of $\CEI$. A couple of striking points:
\begin{itemize}

\item \textbf{the respective role of the two factors}: within this citation window, the influence of growth is small, the $\CEI$ is mostly shaped by the asymmetry of exchanges. However, the dominant role of balance increases with the level of aggregation. In average over years, the ratio is about $8.2$ at the category level, $10.6$ at the sub-discipline level, $14.1$ at the discipline level. With respect to the reduction of all variances in the aggregation process, it appears that the smoothing effect is stronger for differential growth than for balances. This is not unexpected: the status of exporter of knowledge in generic domains (fundamental biology) tends to persist over levels. Conversely, the domains in medical research tend to remain importers of knowledge whatever the level of aggregation, from specialties through "medical research" as a large discipline. This depends on the properties of the network at various scales. We limited ourselves to the three levels mentioned, but a step further if we were to consider "life sciences" as an ensemble, the balance would largely collapse. In terms of trend, both the dispersion of balance and growth slightly decline, except at the discipline level with a quasi-stability of growth between 2001 and 2006.
\item \textbf{the covariance of growth and balance}: covariance is almost always negative over the period. The negative covariance is still higher in absolute value in the non weighted option: domains' size (volume of publications) does not matter and then small domains gain relative importance, among them emerging ones. There is a clear trend over the period, an increase of covariance which seems to get closer to the zero value, remembering however that the last two years are not strictly comparable to the rest of the series. This trend is watched whatever the level of aggregation.
\item \textbf{the variance of impacts}, in terms of trend, is less affected than the $\CEI$ by the down-trend, since the increase of covariance (from fairly negative to weakly negative, not to mention the last two years with incomplete information) compensates for the reduction of variances of growth and balance.
A correlation analysis was also conducted, in the same line. To conclude, as far as an analysis based on a fixed nomenclature can be trusted, there is no sign of differentiation of growth rates over the period, nor of increasing asymmetry in the system, whatever the scale. 
\end{itemize}

\begin{figure}
\begin{center}
\textit{category level:}\\\includegraphics[scale=0.8]{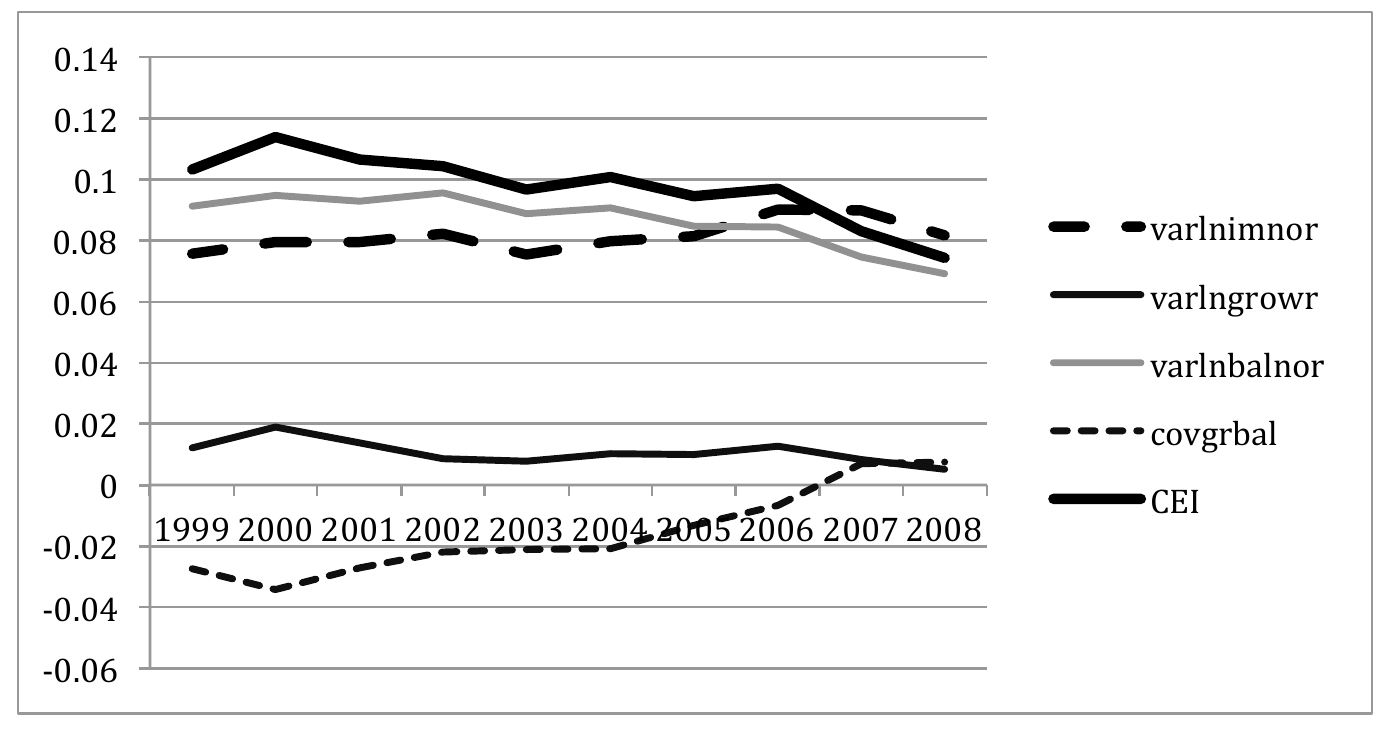}\\
\textit{sub-discipline level:}\\\includegraphics[scale=0.8]{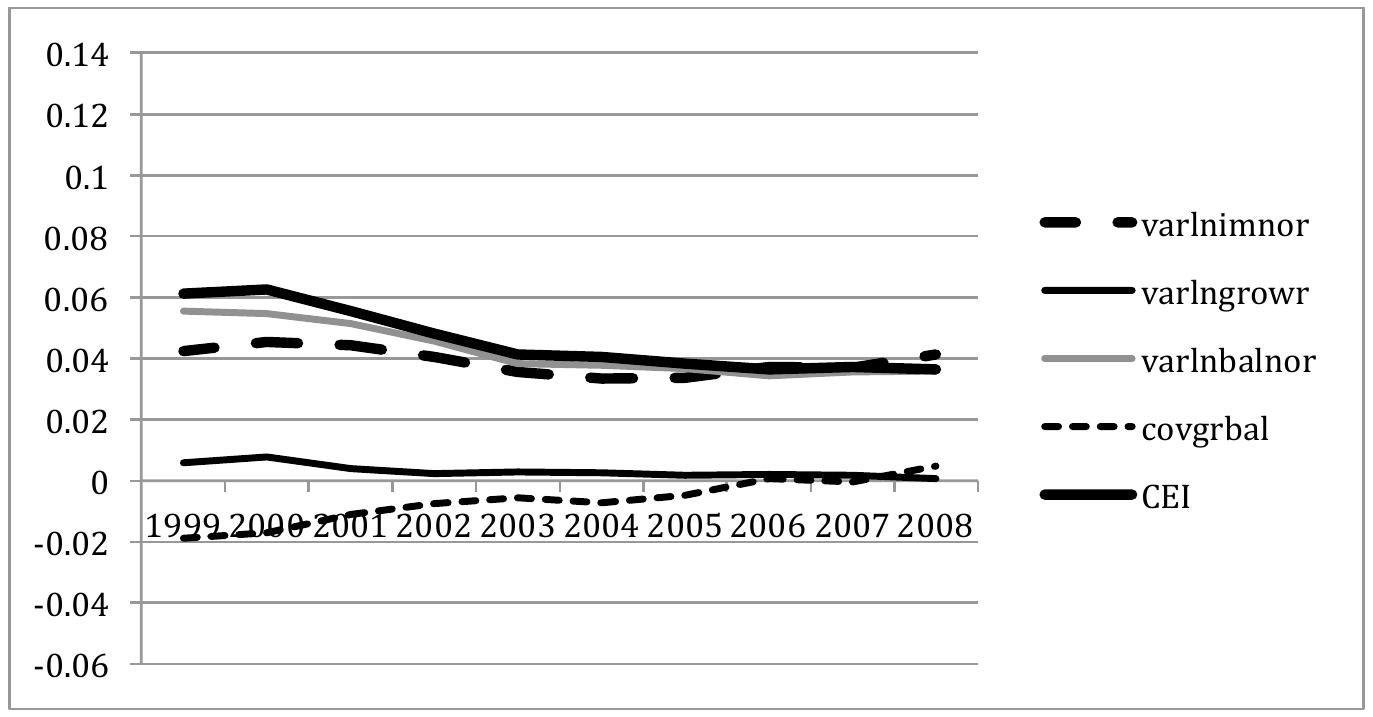} \\
\textit{discipline level:}\\\includegraphics[scale=0.8]{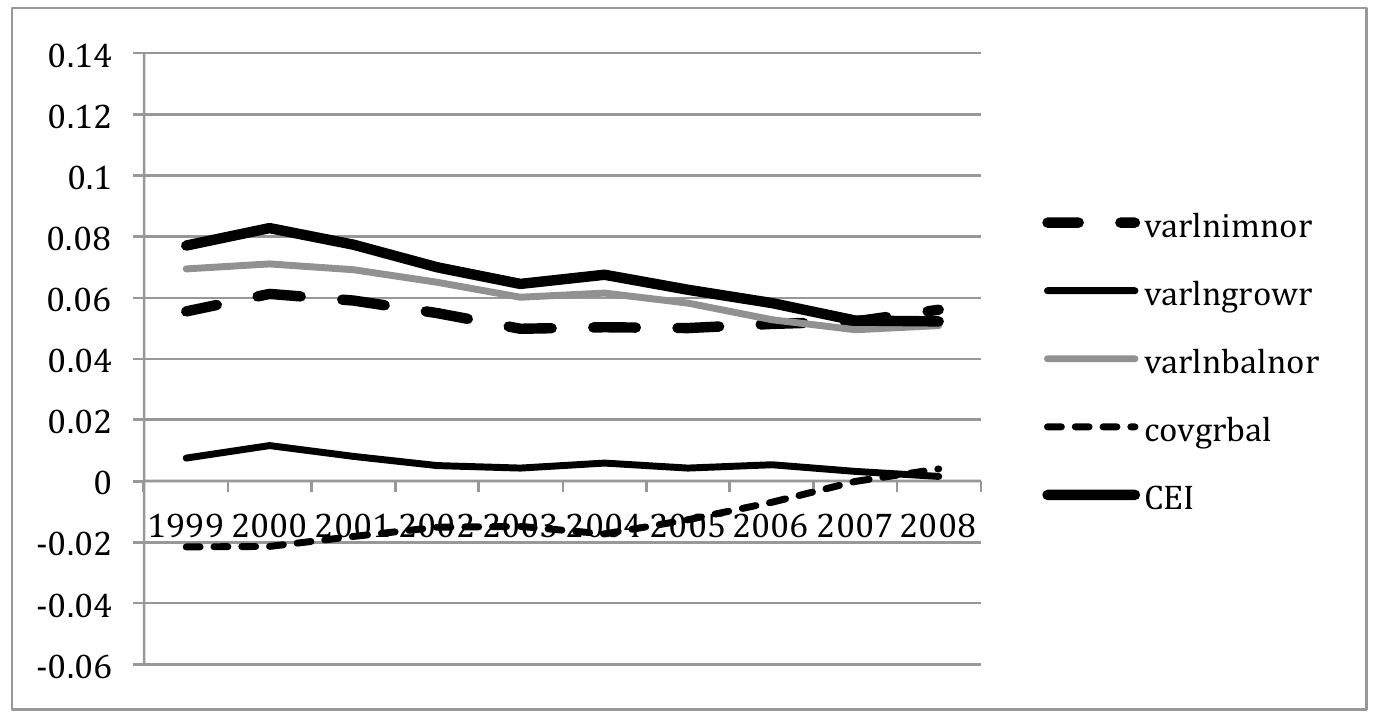} 
\end{center}
\caption{time series of normalized impact and $\CEI$ - variances and covariances at category, sub-discipline and discipline levels (resp. top, middle and bottom)\\
varlnimnor=$\wVar(LI)$: variance of logarithm of normalized impact \\
varlngrowr=$\wVar(LG)$: variance of logarithm of relative growth\\
varlnbalnor=$\wVar(LB)$: variance of logarithm of citation balance\\
covgrbal=$\wCov(LG,LB)$: covariance (logs(growth---balance))\\
$\CEI$ : Change-Exchange Index
}
\label{flows}
\end{figure}

\section{Discussion}
We emphasized that citation impacts corrected for citation propensity identify, at a given domain level, with the product of growth and export-import balance; over all these  domains (science), the variance of those factors are markers of differential growth and of interdisciplinarity in a particular form, the citation dependence. These properties suggest that the variance of impacts one the one hand, and a related index, the Change-Exchange Index, on the other hand, help to partly characterise structural dynamism of the scientific system at a particular level of aggregation. Without scale correcting factor, those measures hold iff all variables involved are defined at the same level of nomenclature (identical level for  the calculation of impact and the normalization). Practical applications of citing-side normalization suppose that a unique level is chosen for averaging the bibliographies lengths (propensity to cite), creating a specific weighted citation network. Normalized impacts can then be calculated for various levels of aggregation.  Therefore, the constraint for establishing the propositions 1, 2 and 3 (the equality of levels) is not satisfied. Interpretation in terms of growth and balance would need the scale correcting factor mentioned above.  

\subsection*{From nomenclature to clusters}
The experiment was conducted on a fixed nomenclature. This is clearly a limitation. Nomenclatures such as databases classification schemes suffer shortcomings: artefacts in the delineation of categories, low reactivity in the short term, sensitivity to national context. These schemes are conservative and let tensions accumulate in the system between two revisions: they may for example keep two sub-domains attached, that a data analysis based on bibliometric networks could consider as having parted, and conversely for merges. An alternative is to rely on those networks, especially citations, with various transformations \citep{kessler1963bibliographic,small1973co,marshakova1973document,Chen:1999vv,boyack2005mapping}, which proved powerful tools for clustering and mapping science. Clustering reduces tensions by adjusting delineation of domain and trading topics. Substituting clusters/ neighborhoods to categories is therefore expected to yield more realistic representations, minimizing artificial exchange flows. Bibliometric clusters (co-authorship, citations, semantic content) enable scholars to track emergence and life-cycle phenomena \citep{scharnhorst2012models, morris2005manifestation, chavalarias2013phylomemetic}. If citation approach is preferred, which is logical in our context, a sensible objective is to reach bipartite clusters encompassing both cited and citing items in close relation. The choice of symmetrical metrics (citing<-->cited) seems preferable for building clusters, avoiding mixing areas with asymmetrical positions in the flows of knowledge. 

A challenge of network-based clustering is the loss of coverage: in nomenclature schemes or classification based on editorial entities (journals), any citable article is classified, whether cited or not; any citing article is classified, whether its references are "active" (falling into the citation window) or not. For example, the exercise of "audience factor" \citep{zitt2008modifying} based on entire journals on both sides, escapes this integrity issue, whereas finer granularity exercises (see SNIP but also \cite{moed2010measuring}) have to cope with it. There are various ways – easier for citing than for citable articles – to circumvent the problem, by relaxing the citation window, modifying the construction of neighbourhood, introducing a correcting factor (SNIP 2, Waltman \& van Eck, 2012).

The application of bibliometric clustering to the questions addressed here is promising. It would be appealing to confirm the slant suggested by the empirical findings, a relative down-trend in differentiation of growth rates and asymmetry of domains' balances.

It should be stressed that unlike the conventional cited-side normalization, the citing-side approach does not aim at a complete normalization. Usual quality tests assessing the performance of the various methods on the ground of the total reduction of variability are inappropriate in the present context. By limiting itself to the correction of propensity to cite, the citing-side approach reveals fruitful. We have focused in this paper especially on these uncontrolled factors.

\subsection*{Normalized Impact, Change-Exchange Index and dynamics of science}
Further research is needed to explore the various aspects of these measures. A first question is the effect of the citation window's length. A more general issue is the linkage between macro and micro-models. The relationship between growth and balance along the typical life cycles of scientific domains is appealing. We gave empirical evidence that growth rate and balance are negatively correlated. The equilibrium between values of growth and balance variance on the one hand– with respect to citation windows -  their covariance on the other hand are linked to features of local structure and dynamics. The sign of covariance, all things equal, may change over different phases in a domain's life-cycle. Typically negative in emergence phases, it may become positive in the central phases, especially if endogenous growth is echoed by external diffusion in the network overcoming domains' borders. A challenge is to connect model of life-cycle of areas, preferably delineated by citation-based clustering, with various mechanisms of networks dynamics (Powell et al. 2005), among them preferential attachment \citep{de1963little,Jeong:2007wi, eom2011characterizing}.
 
The present approach only addresses aggregate phenomena. Balances at the domain level express a particular aspect of inter-disciplinarity, the asymmetrical linkages: domains equalizing exports and imports of knowledge will tend to reduce the dispersion. Many dimensions of citation networks are not accounted for. Diversity, essential for understanding the science structure and dynamics, is not directly adressed. It should also be stressed that only relative changes were addressed here, through relative variables. The absolute growth or the average impact over science is corrected for, in contrast with long-range analyses in the wake of Price (1963) which focus on volumes of publications and citations (see for example \cite{lariviere2007long} in their study of aging). 

To conclude, citing-side approach opens a new perspective for the analysis of knowledge flows, insofar as they can be sketched by citation networks. A citing-side weighting is a promising solution for addressing diversity and interdisciplinary studies \citep{zitt2011behind,Rafols:2012hv}, with a significant improvement over gross flows analysis (e.g. \cite{rinia2002measuring}). Here, at a macro-level, we have shown that basic relations connect the novel normalized impact and a derived measure, the $\CEI$, to important features of dynamism and structure of science. The relation with the parallel and powerful "influence weighting" pioneered by \cite{narin1976evaluative} with iterative weighting of citation sources, that has known a revival in the last decade \citep{palacios2004measurement,Bergstrom:2007uf} is also appealing.

\paragraph{Acknowledgements:}
The authors thank OST (S. Ramanana and G. Filliatreau) for help in providing aggregated data and bases of nomenclature. Modifications are the responsibility of the authors. They also thank E. Bassecoulard for her assistance.


\section*{Appendix 1}
The basic propositions are established by considering that the domains are used as the basis of definition for all variables involved: relative growth; balance of citation exchanges; relative number of references; and the resulting relative impact. "Relative" values are understood as the ratio to the value for all science. The basic equations hold for any sensible level of granularity where it makes sense to align citing and citable literature. For convenience, the empirical illustrations are based three levels (subject category; sub-discipline; discipline) in a fixed nomenclature, albeit the spirit of citing-side normalization is nomenclature-free, in contrast with classical normalization.

For a given domain $A$ of science $S$ ($A \in S$), we distinguish the set of articles pertaining to $A$  according to their publication  period $T_0$ and $T_1$ (respectively $A^{T_0}$ and $A^{T_1}$). For the sake of clarity we will assume that the two time periods have the same length. We can define the matrix $C$ which summarizes every citations pertaining to  $A$, \emph{e.g.} every citation links from articles in $A$ or pointing to articles in $A$. Obviously $C$ is binary and asymmetric: for two articles $A$ and $j$, $C(i,j) = 1 \text{ iff } i \text{ cites } j$.  For the sake of clarity, we assume that only articles written during $T_0$ ($A^{T_0}$) are cited and that these  citations are emitted  by articles produced during period $T_1$.  It can also be useful to interpret the citation matrix as a directed bipartite graph: $G=(S,C)$ featuring a set of articles $i \in S$ tagged according to their domain of science and organized in two sets according to their publication date connected by the set of citations links: $C$.  The total number of citations received by a publication is then simply given by its in-degree in $G$.

   Those citations can be aggregated at the domain level: incoming and out-going citations at the domain level $A$ will be respectively denoted: $\Phi_\rightarrowtail(A) = \sum_{i \in S,~ j \in A}C(i,j)$ and $\Phi_\leftarrowtail(A)= \sum_{i\in A,~ j \in S }C(i,j)$. 

We define the growth rate $\rho(A)$ of the domain $A$ as the publication number growth rate between the two successive periods (periods of same length, or appropriate annual averages on the citing, respectively the cited period) : $\rho(A)=\frac{|A^{T_1}|}{|A^{T_0}|}$.
{We can also define $\hat{\rho}(A)$, $A$'s \emph{relative growth rate}  with respect with the general growth rate of science by computing the ratio between its growth rate between the two successive time periods with the global growth rate assessed at the whole science scale: 

$$
 \label{growth}
\hat{\rho}(A) = \frac{|A^{T_1}|/|A^{T_0}|}{|S^{T_1}|/|S^{T_0}|}
$$

It should be kept in mind that the growth rate depends on the citing period with respect to the cited period, just as the balances defined below, the relative length of bibliography when needed, and the final relative impact.

The impact of a domain $A$ is defined by the average number of incoming citations per citable articles in $A$:
$$ \label{impactA}
I(A) = \frac{\Phi_{\leftarrow}(A)} {|A^{T_0}|}
$$
The relation $I(S) = \frac{\Phi_{\leftarrow}(S)} {|S^{T_0}|}$ can be written $I(S) = \frac{\kappa(S^{T_1})|S^{T_1}|} { |S^{T_0}|}$ as the total incoming citation flows can be written as the total number of citing papers times the average number of references in citing  articles $\kappa(S^{T_1})$. It then follows that :
$$
 \label{impactS}
I(S) = \kappa(S^{T_1}) \rho(S)
$$

We can also detail the citation in-flow according to the sources, which yield to this expression of  the impact of a domain $A$:

\begin{equation}
 \label{impact}
I(A) =  \frac{\Phi_{\leftarrow}(A)}{\Phi_{\rightarrow}(A)} \frac{\Phi_{\rightarrow}(A)}{|A^{T_0}|}
\end{equation}

We then define the balance ratio, which compares the total inflow with the total outflow of $A$, such as :

$$
B(A) = \frac{\Phi_\leftarrowtail(A)}{\Phi_\rightarrowtail(A)}
$$

%

Combining the two previous equations, the impact of a domain $A$  can be described as:
$$I(A) =  B(A)\frac{\Phi_{\rightarrow}(A)}{|A^{T_0}|}$$

By definition, $\Phi_{\rightarrow}(A)=\kappa(A^{T_1})|A^{T_1}|$, the global equation then rewrites:

$$I(A) =  B(A)\kappa(A)\rho(A)$$

Defining the relative impact $\hat{I}(A)$ of $A$ as the absolute impact $I(A)$ divided by the absolute impact of all science $I(S)$, it comes:
$$\label{global}
\hat{I}^g(A) =  B^g(A)\hat{\rho}(A)
$$

where $\hat{\kappa}(A)=\frac{\kappa(A)}{\kappa(S)}$  is the average number of out-going citation in $A$ normalized with respect to $S$, that is the relative length of bibliography.

This gross relative impact then depends on three components linked to dynamic aspects (relative growth $\hat{\rho}$ and exchanges B) and the variations of citation habits $\hat{\kappa}$. Interpretation in terms of dynamics of science could be possible based on a raw citation flows count, differential growth and exchanges being elements of changes in the system but they would be blurred by citation habits.

The purpose of citing side normalization of citation flows is to get rid of those variations which can be quite large (between one and two orders of magnitude at the "subject category" level). The citing-side normalization neutralizes the factor of citation habits $\hat{\kappa}=1$ and makes comparison possible on the whole system.

From the original citation network, we can derive  $G^g=(S,C^g)$  where citations links $C^g$ are normalized with respect to the average propensity to cite of each domain. Each citation coming from a publication in $A$ is assigned a weight. This procedure provides more weight to citations stemming from domains producing fewer citations on average. The edges of $C^g$ coming from publications in are then weighted according to the formula:

$$C^g(i,j) = C(i,j)  \frac{w^g(I)}{w^g(S)}$$
where $w^g(I)$ and $w^g(S)$ are the average number of citations produced by publications published in  $I$ or in $S$: $w^g(I) = \frac{\sum_{i\in I,j}C(i,j)}{|A^{T_1}|}$ and $w^g(S)=  \frac{\sum_{i,j}C(i,j)}{|S^{T_1}|}$.

The same argument regarding citation flows and related impacts holds with this new normalized definition of the citation matrix, such that the general equation is simplified:
$$\label{global}
\hat{I}^g(A) =  B^g(A)\hat{\rho}(A)
$$

\clearpage
\section*{Appendix 2}
\begin{scriptsize}
\begin{center}
\begin{longtable}{|p{2.8cm}|p{6.cm}|p{6cm}|}
\hline & & \\
DISCIPLINES & Sub-Disciplines & Categories\\ 
  & & \\ \hline & & \\
  APPLIED BIOLOGY & Agriculture, Plant Sciences & Agricultural Economics \& Policy\\ 
  &   & Agronomy\\ 
  &   & Plant Sciences\\ 
  &   & Forestry\\ 
  &   & Horticulture\\ 
  &   & Materials Science, Paper \& Wood\\ 
  &   & Materials Science, Textiles\\ 
  &   & Soil Science\\ 
  & Agro-Industry, Food-Industry  & Agriculture, Dairy \& Animal Sciences\\ 
  &   & Agricultural Engineering\\ 
  &   & Agriculture, Multidisciplinary\\ 
  &   & Food Science \& Technology\\ 
  & Biology, Miscellaneous & Biology, Miscellaneous\\ 
  &   & Mycology\\ 
  & Ecology, Marine Biology & Biodiversity, Conservation\\ 
  &   & Ecology\\ 
  &   & Engineering, Marine\\ 
  &   & Entomology\\ 
  &   & Fisheries\\ 
  &   & Marine \& Freshwater Biology\\ 
  &   & Ornithology\\ 
  &   & Zoology\\ 
    & & \\ \hline & & \\
CHEMISTRY & Analytical Chemistry & Biochemical Research Methods\\ 
  &   & Chemistry, Analytical\\ 
  & General Chemistry & Chemistry, Multidisciplinary\\ 
  & Materials \& Polymer Science & Materials Science, Ceramics\\ 
  &   & Materials Science, Multidisciplinary\\ 
  &   & Metallurgy \& Metallurgical Engineering\\ 
  &   & Materials Science, Characterization\\ 
  &   & Materials Science, Composites\\ 
  &   & Polymer Science\\ 
  & Chemistry, Organic, Mineral, Nuclear  & Chemistry, Applied\\ 
  &   & Chemistry, Inorganic \& Nuclear\\ 
  &   & Chemistry, Organic\\ 
  & Chemistry, Physical & Chemistry, Physical\\ 
  &   & Crystallography\\ 
  &   & Electrochemistry\\ 
  &   & Materials Science, Coatings \&Films\\ 
    & & \\ \hline & & \\
EARTH \& SPACE SCIENCES & Astronomy, Astrophysics & Astronomy \& Astrophysics\\ 
  & Environment & Engineering, Environmental\\ 
  &   & Environmental Sciences\\ 
  &   & Limnology\\ 
  &   & Water Resources\\ 
  & Geosciences & Engineering, Aerospace\\ 
  &   & Geochemistry \& Geophysics\\ 
  &   & Engineering, Geological\\ 
  &   & Geography, Physical\\ 
  &   & Geology\\ 
  &   & Geosciences, Multidisciplinary\\ 
  &   & Meteorology \& Atmospheric Sciences\\ 
  &   & Mineralogy\\ 
  &   & Oceanography\\ 
  &   & Remote Sensing\\ 
  &   & Paleontology\\ 
  &   & Imaging Science \& Photographic Technology\\ 
    & & \\ \hline & & \\
ENGINEERING \& COMPUTER SCIENCES & Civil Engineering, Mining & Construction \& Building Technology\\ 
  &   & Engineering, Civil\\ 
  &   & Engineering, Ocean\\ 
  &   & Transportation Science \& Technology\\ 
  &   & Mining \& Mineral Processing\\ 
  & Energy, Chemical \& Industrial Engineering & Acoustics\\ 
  &   & Thermodynamics\\ 
  &   & Energy \& Fuels\\ 
  &   & Engineering, Multidisciplinary\\ 
  &   & Engineering, Chemical\\ 
  &   & Engineering, Petroleum\\ 
  &   & Engineering, Mechanical\\ 
  &   & Mechanics\\ 
  & ICST,Robotics, Artificial Intelligence & Automation \& Control Systems\\ 
  &   & Computer Science, Artificial Intelligence\\ 
  &   & Computer Science, Cybernetics\\ 
  &   & Engineering, Industrial\\ 
  &   & Engineering, Manufacturing\\ 
  &   & Operations Research \& Managemement Science\\ 
  &   & Robotics\\ 
  & ICST, Electrical \& Electronic Engineering & Engineering, Electrical \& Electronic\\ 
  & ICST, Informatics And Communication & Computer Science, Hardware \& Architecture\\ 
  &   & Computer Science, Information\\ 
  &   & Computer Science, Interdisciplinary\\ 
  &   & Computer Science, Software Engineering\\ 
  &   & Computer Science, Theory \& Methods\\ 
  &   & Telecommunications\\ 
    & & \\ \hline & & \\
FUNDAMENTAL BIOLOGY & Biochemistry & Biochemistry \& Molecular Biology\\ 
  &   & Biology\\ 
  &   & Biophysics\\ 
  &   & Cell Biology\\ 
  &   & Physiology\\ 
  & Bioengineering & Health Care Sciences \& Services\\ 
  &   & Engineering, Biomedical\\ 
  &   & Mathematical \& Computational Biology\\ 
  &   & Medical Informatics\\ 
  &   & Materials Science, Biomaterials\\ 
  &   & Neuroimaging\\ 
  &   & Radiology, Nuclear Medicine \&Imaging\\ 
  & Biotechnology, Genetics & Cell \& Tissue Engineering\\ 
  &   & Biotechnology \& Applied Microbiology\\ 
  &   & Genetics \& Heredity\\ 
  & Endocrinology & Endocrinology \& Metabolism\\ 
  &   & Nutrition \& Dietetics\\ 
  & Neurosciences & Behavioral Sciences\\ 
  &   & Neurosciences\\ 
  & Reproduction, Developmental Biology I & Anatomy \& Morphology\\ 
  &   & Evolutionary Biology\\ 
  &   & Developmental Biology\\ 
  &   & Microscopy\\ 
  &   & Reproductive Biology\\ 
    & & \\ \hline & & \\
HUMAN SCIENCES & Geography, Demography, Ethnography & Area Studies\\ 
&   & Demography\\ 
  &   & Ethnic Studies\\ 
  &   & Folklore\\ 
  &   & Geography\\ 
  & History, Archeology & Anthropology\\ 
  &   & Archaeology\\ 
  &   & History \& Philosophy Of Sciences\\ 
  &   & History Of Social Sciences\\ 
  & HUMAN/Linguistics \& Communication & Communication\\ 
  &   & Language \& Linguistics\\ 
  & Literature, Philosophy) & Ethics\\ 
  &   & Linguistics\\ 
  &   & Philosophy\\ 
  & Psychology & Psychology, Biological\\ 
  &   & Psychology, Clinical\\ 
  &   & Psychology, Educational\\ 
  &   & Psychology, Developmental\\ 
  &   & Psychology, Applied\\ 
  &   & Psychology, Multidisciplinary\\ 
  &   & Psychology, Psychoanalysis\\ 
  &   & Psychology, Mathematical\\ 
  &   & Psychology, Experimental\\ 
  &   & Psychology, Social\\ 
    & & \\ \hline & & \\
MATHEMATICS & Mathematics & Mathematics, Applied\\ 
  &   & Mathematics, Interdisciplinary\\ 
  &   & Mathematics\\ 
  &   & Statistics \& Probability\\ 
    & & \\ \hline & & \\
MEDICAL RESEARCH & Oncology & Oncology\\ 
  & Cardiology, Pneumology & Cardiac \& Cardiovascular Systems\\ 
  &   & Critical Care Medicine\\ 
  &   & Emergency Medicine\\ 
  &   & Hematology\\ 
  &   & Respiratory System\\ 
  &   & Peripheral Vascular Disease\\ 
  & Medicine, Miscellaneous & Anesthesiology\\ 
  &   & Dentistry, Oral Surgery \& Medicine\\ 
  &   & Dermatology\\ 
  &   & Medical Laboratory Technology\\ 
  &   & Medicine, General \& Internal\\ 
  &   & Medicine, Research \& Experimental\\ 
  &   & Ophthalmology\\ 
  &   & Otorhinolaryngology\\ 
  &   & Pathology\\ 
  &   & Rheumatology\\ 
  &   & Veterinary Sciences\\ 
    & & \\ \hline & & \\
MEDICAL RESEARCH & Public Health, Miscellaneous & Substance Abuse\\ 
  &   & Geriatrics \& Gerontology\\ 
  &   & Public, Environmental \& Occupational Health\\ 
  &   & Integrative \& Complementary Medicine\\ 
  &   & Medical Ethics\\ 
  &   & Medicine, Legal\\ 
  &   & Nursing\\ 
  &   & Orthopedics\\ 
  &   & Parasitology\\ 
  &   & Pediatrics\\ 
  &   & Rehabilitation\\ 
  &   & Sport Sciences\\ 
  &   & Tropical Medicine\\ 
  & Microbiolog \& Virology, Immunology & Allergy\\ 
  &   & Immunology\\ 
  &   & Infectious Diseases\\ 
  &   & Microbiology\\ 
  &   & Virology\\ 
  & Clinical Neurosciences & Clinical Neurology\\ 
  &   & Psychiatry\\ 
  &   & Psychology\\ 
  & Pharmacy \& Toxicology & Chemistry, Medicinal\\ 
  &   & Pharmacology \& Pharmacy\\ 
  &   & Toxicology\\ 
  & Reproduction, Developmental Biology II & Andrology\\ 
  &   & Obstetrics \& Gynecology\\ 
  & Surgery, Gastroenterology, Urology & Gastroenterology \& Hepatology\\ 
  &   & Surgery\\ 
  &   & Transplantation\\ 
  &   & Urology \& Nephrology\\ 
    & & \\ \hline & & \\
MULTI-DISCIPLINARY & Multidisciplinary & Education, Scientific Disciplines\\ 
  &   & Multidisciplinary Sciences\\ 
  & Nanosciences \& Technology & Nanoscience \& Nanotechnology\\ 
    & & \\ \hline & & \\
PHYSICS & Physics, Condensed Matter & Physics, Applied\\ 
  &   & Physics, Condensed Matter\\ 
  & General Physics & Physics, Fluids \& Plasmas\\ 
  &   & Physics, Multidisciplinary\\ 
  &   & Physics, Mathematical\\ 
  & Particles \& Nuclear Physics & Instruments \& Instrumentation\\ 
  &   & Nuclear Science \& Technology\\ 
  &   & Optics\\ 
  &   & Physics, Atomic- Molecular \& Chemical\\ 
  &   & Physics, Nuclear\\ 
  &   & Physics, Particles \& Fields\\ 
  &   & Spectroscopy\\ 
    & & \\ \hline & & \\
SOCIAL SCIENCES & Economics & Economics\\ 
  & Education & Education \& Educational Research\\ 
  &   & Education, Special\\ 
  & Public Health, Social & Ergonomics\\ 
  &   & Gerontology\\ 
  &   & Health Policy \& Services\\ 
  &   & Social Sciences, Biomedical\\ 
  &   & Social Work\\ 
  & Information \& Library Science & Information Science \& Library Science\\ 
  & Law & Criminology \& Penology\\ 
  &   & Law\\ 
  & Management; Business, Finance & Business\\ 
  &   & Business, Finance\\ 
  &   & Management\\ 
  & Political Sciences  & International Relations\\ 
  &   & Planning \& Development\\ 
  &   & Political Science\\ 
  &   & Public Administration\\ 
  & Public Policies & Environmental Studies\\ 
  &   & Social Issues\\ 
  &   & Transportation\\ 
  &   & Urban Studies\\ 
  & Sociology & Family Studies\\ 
  &   & Industrial Relations \& Labor\\ 
  &   & Social Sciences, Mathematical\\ 
  &   & Social Sciences, Interdisciplinary\\ 
  &   & Sociology\\ 
  &   & Women's Studies\\ 
  & & \\ \hline
  \caption{Source:  WoS for the category level, OST for the initial aggregations into sub-disciplines and disciplines (G.Filliatreau, S. Ramanana, M. Zitt). The original scheme was modified by the authors to get a hierarchical structure. NB: the nomenclature above only presents the selected categories, some of them not present throughout the period. In addition, only couples year-category with the category present both on the citable and cited side were retained. Because of scarce average references-citations, the following categories were discarded from the study: ARCHITECTURE, ART,HUMANITIES-MULTIDISCIPLINARY,CULTURAL STUDIES, LITERATURE CLASSICS, DANCE, CINEMA-RADIO-TELEVISION, FOLKLORE, PRIMARY HEALTH CARE, HISTORY, LEISURE-SPORT, ASIAN STUDIES, LITERARY THEORY \& CRITICISM, LITERARY REVIEWS, LITERATURE, LITERATURE-AFRICAN-AUSTRALIAN, LITERATURE-AMERICAN, LITERATURE-BRITISH ISLES, LITERATURE-GERMAN-DUTCH-SCANDINAVIAN, LITERATURE- ROMANE, LITERATURE-SLAVIC, MEDIEVAL \& RENAISSANCE STUDIES, MUSIC, POETRY, THEATER, RELIGION
}
\end{longtable}

\end{center}
\end{scriptsize}



\begin{thebibliography}{}

\bibitem[\protect\citename{Bergstrom, }2007]{Bergstrom:2007uf}
Bergstrom, Carl~T. 2007.
\newblock {Eigenfactor Measuring the value and prestige of scholarly journals}.
\newblock {\em College {\&} Research Libraries News}, {\bf 68}(5).

\bibitem[\protect\citename{Boyack {\em et~al.\ }\relax,
  }2005]{boyack2005mapping}
Boyack, Kevin~W, Klavans, Richard, \& B~orner, Katy. 2005.
\newblock {Mapping the backbone of science}.
\newblock {\em Scientometrics}, {\bf 64}(3), 351--374.

\bibitem[\protect\citename{Chavalarias \& Cointet,
  }2013]{chavalarias2013phylomemetic}
Chavalarias, David, \& Cointet, Jean-Philippe. 2013.
\newblock Phylomemetic Patterns in Science Evolution - The Rise and Fall of
  Scientific Fields.
\newblock {\em PLOS ONE}, {\bf 8}(2), e54847.

\bibitem[\protect\citename{Chen, }1999]{Chen:1999vv}
Chen, C. 1999.
\newblock {Visualising semantic spaces and author co-citation networks in
  digital libraries}.
\newblock {\em Information Processing {\&} Management}, {\bf 35}(3), 401--420.

\bibitem[\protect\citename{Czapski, }1997]{Czapski:1997jh}
Czapski, G. 1997.
\newblock {The use of deciles of the citation impact to evaluate different
  fields of research in Israel}.
\newblock {\em Scientometrics}, {\bf 40}(3), 437--443.

\bibitem[\protect\citename{de~Solla~Price, }1965]{deSollaPrice:1965vs}
de~Solla~Price, D. 1965.
\newblock {Networks of Scientific Papers}.
\newblock {\em Science (New York, NY)}.

\bibitem[\protect\citename{de~Solla~Price, }1963]{de1963little}
de~Solla~Price, D~J. 1963.
\newblock {\em {Little science, big science}}.
\newblock George B. Pegram lectures.
\newblock Columbia University Press.

\bibitem[\protect\citename{Eom \& Fortunato, }2011]{eom2011characterizing}
Eom, Young-Ho, \& Fortunato, Santo. 2011.
\newblock Characterizing and modeling citation dynamics.
\newblock {\em PloS one}, {\bf 6}(9), e24926.

\bibitem[\protect\citename{Garfield, }1955]{Garfield:1955tm}
Garfield, Eugene. 1955.
\newblock {Citation Indexes for Science - A New Dimension in Documentation
  through Associationof Ideas}.
\newblock {\em Science (New York, NY)}, July, 108--111.

\bibitem[\protect\citename{Garfield, }2006]{Garfield:2006gm}
Garfield, Eugene. 2006.
\newblock {The history and meaning of the journal impact factor.}
\newblock {\em JAMA : the journal of the American Medical Association}, {\bf
  295}(1), 90--93.

\bibitem[\protect\citename{Garfield \& Merton, }1979]{Garfield:1979uz}
Garfield, Eugene, \& Merton, Robert~K. 1979.
\newblock {Citation indexing-its theory and application in science, technology,
  and humanities, Eugene Garfield ; foreword by Robert K. Merton.}
\newblock {\em New York : John Wiley}.

\bibitem[\protect\citename{Gl{\"a}nzel {\em et~al.\ }\relax,
  }2011]{Glanzel:2011df}
Gl{\"a}nzel, Wolfgang, Schubert, Andras, Thijs, Bart, \& Debackere, Koenraad.
  2011.
\newblock {A priori vs. a posteriori normalisation of citation indicators. The
  case of journal ranking}.
\newblock {\em Scientometrics}, {\bf 87}(2), 415--424.

\bibitem[\protect\citename{Jeong {\em et~al.\ }\relax, }2007]{Jeong:2007wi}
Jeong, H, N{\'e}da, Z, \& Barabasi, A-L. 2007.
\newblock {Measuring preferential attachment in evolving networks}.
\newblock {\em EPL (Europhysics Letters)}, {\bf 61}(4), 567.

\bibitem[\protect\citename{Kessler, }1963]{kessler1963bibliographic}
Kessler, M.M. 1963.
\newblock Bibliographic coupling between scientific papers.
\newblock {\em American documentation}, {\bf 14}(1), 10--25.

\bibitem[\protect\citename{Larivi{\`e}re {\em et~al.\ }\relax,
  }2007]{lariviere2007long}
Larivi{\`e}re, Vincent, Archambault, {\'E}ric, \& Gingras, Yves. 2007.
\newblock Long-term variations in the aging of scientific literature: From
  exponential growth to steady-state science (1900--2004).
\newblock {\em Journal of the American Society for Information Science and
  technology}, {\bf 59}(2), 288--296.

\bibitem[\protect\citename{Leydesdorff \& Opthof, }2010]{Leydesdorff:2010wb}
Leydesdorff, L, \& Opthof, T. 2010.
\newblock {Scopus's source normalized impact per paper (SNIP) versus a journal
  impact factor based on fractional counting of citations}.
\newblock {\em Journal of the American Society for Information Science and
  Technology}, {\bf 61}(11), 2365--2369.

\bibitem[\protect\citename{Marshakova, }1973]{marshakova1973document}
Marshakova, I~V. 1973.
\newblock {Document coupling system based on references taken from Science
  Citation Index}.
\newblock {\em Russia, Nauchno-Teknicheskaya Informatsiya, Ser}, {\bf 2}.

\bibitem[\protect\citename{Moed, }2010]{moed2010measuring}
Moed, H~F. 2010.
\newblock {Measuring contextual citation impact of scientific journals}.
\newblock {\em Journal of Informetrics}, {\bf 4}(3), 265--277.

\bibitem[\protect\citename{Morris, }2005]{morris2005manifestation}
Morris, Steven~A. 2005.
\newblock Manifestation of emerging specialties in journal literature: A growth
  model of papers, references, exemplars, bibliographic coupling, cocitation,
  and clustering coefficient distribution.
\newblock {\em Journal of the American Society for Information Science and
  Technology}, {\bf 56}(12), 1250--1273.

\bibitem[\protect\citename{Murugesan \& Moravcsik,
  }1978]{murugesan2007variation}
Murugesan, P, \& Moravcsik, M~J. 1978.
\newblock {Variation of the nature of citation measures with journals and
  scientific specialties}.
\newblock {\em Journal of the American Society for Information Science and
  Technology}, {\bf 29}(3), 141--147.

\bibitem[\protect\citename{Narin, }1976]{narin1976evaluative}
Narin, F. 1976.
\newblock {\em {Evaluative bibliometrics: The use of publication and citation
  analysis in the evaluation of scientific activity}}.
\newblock Computer Horizons Washington, D. C.

\bibitem[\protect\citename{Palacios-Huerta \& Volij,
  }2004]{palacios2004measurement}
Palacios-Huerta, I, \& Volij, O. 2004.
\newblock {The measurement of intellectual influence}.
\newblock {\em Econometrica}, {\bf 72}(3), 963--977.

\bibitem[\protect\citename{Radicchi {\em et~al.\ }\relax,
  }2008]{radicchi2008universality}
Radicchi, F, Fortunato, S, \& Castellano, C. 2008.
\newblock {Universality of citation distributions: Toward an objective measure
  of scientific impact}.
\newblock {\em Proceedings of the National Academy of Sciences}, {\bf 105}(45),
  17268--17272.

\bibitem[\protect\citename{Rafols {\em et~al.\ }\relax, }2012]{Rafols:2012hv}
Rafols, Ismael, Leydesdorff, Loet, O'Hare, Alice, Nightingale, Paul, \&
  Stirling, Andy. 2012.
\newblock {How journal rankings can suppress interdisciplinary research: A
  comparison between Innovation Studies and Business {\&} Management}.
\newblock {\em Research Policy}, {\bf 41}(7), 1262--1282.

\bibitem[\protect\citename{Redner, }2005]{redn:cita}
Redner, S. 2005.
\newblock {Citation Statistics from 110 Years of Physical Review}.
\newblock {\em Physics Today}, {\bf 58}, 49--54.

\bibitem[\protect\citename{Rinia {\em et~al.\ }\relax,
  }2002]{rinia2002measuring}
Rinia, E~J, Van~Leeuwen, T~N, Bruins, E E~W, Van~Vuren, H~G, \& Van~Raan, A
  F~J. 2002.
\newblock {Measuring knowledge transfer between fields of science}.
\newblock {\em Scientometrics}, {\bf 54}(3), 347--362.

\bibitem[\protect\citename{Scharnhorst {\em et~al.\ }\relax,
  }2012]{scharnhorst2012models}
Scharnhorst, Andrea, B{\"o}rner, Katy, \& van~den Besselaar, Peter. 2012.
\newblock {\em Models of Science Dynamics: Encounters Between Complexity Theory
  and Information Sciences}.
\newblock Springer.

\bibitem[\protect\citename{Schubert \& Braun, }1986]{schubert1986relative}
Schubert, A, \& Braun, T. 1986.
\newblock {Relative indicators and relational charts for comparative assessment
  of publication output and citation impact}.
\newblock {\em Scientometrics}, {\bf 9}(5), 281--291.

\bibitem[\protect\citename{Sen, }1992]{sen1992normalised}
Sen, B~K. 1992.
\newblock {Normalised impact factor}.
\newblock {\em Journal of Documentation}, {\bf 48}(3), 318--325.

\bibitem[\protect\citename{Small, }1973]{small1973co}
Small, H. 1973.
\newblock {Co-citation in the scientific literature: A new measure of the
  relationship between two documents}.
\newblock {\em Journal of the American Society for Information Science and
  Technology}, {\bf 24}(4), 265--269.

\bibitem[\protect\citename{Small \& Sweeney, }1985]{small1985clustering}
Small, H., \& Sweeney, E. 1985.
\newblock {Clustering the science citation index{\textregistered} using
  co-citations}.
\newblock {\em Scientometrics}, {\bf 7}(3), 391--409.

\bibitem[\protect\citename{Vinkler, }2002]{vinkler2002subfield}
Vinkler, P. 2002.
\newblock {Subfield problems in applying the Garfield (Impact) Factors in
  practice}.
\newblock {\em Scientometrics}, {\bf 53}(2), 267--279.

\bibitem[\protect\citename{Waltman \& van Eck, }2012]{waltman2012inconsistency}
Waltman, L., \& van Eck, N.J. 2012.
\newblock The inconsistency of the h-index.
\newblock {\em Journal of the American Society for Information Science and
  Technology}.

\bibitem[\protect\citename{Yitzhaki, }2003]{yitzhaki2003gini}
Yitzhaki, S. 2003.
\newblock {Gini's mean difference: A superior measure of variability for
  non-normal distributions}.
\newblock {\em Metron}, {\bf 61}(2), 285--316.

\bibitem[\protect\citename{Zitt, }2011]{zitt2011behind}
Zitt, M. 2011.
\newblock {Behind citing-side normalization of citations: some properties of
  the journal impact factor}.
\newblock {\em Scientometrics}, {\bf 89}(1), 329--344.

\bibitem[\protect\citename{Zitt \& Small, }2008]{zitt2008modifying}
Zitt, M, \& Small, H. 2008.
\newblock {Modifying the journal impact factor by fractional citation
  weighting: The audience factor}.
\newblock {\em Journal of the American Society for Information Science and
  Technology}, {\bf 59}(11), 1856--1860.

\bibitem[\protect\citename{Zitt {\em et~al.\ }\relax,
  }2005]{zitt2005relativity}
Zitt, M, Ramanana-Rahary, S, \& Bassecoulard, E. 2005.
\newblock {Relativity of citation performance and excellence measures: From
  cross-field to cross-scale effects of field-normalisation}.
\newblock {\em Scientometrics}, {\bf 63}(2), 373--401.

\end{thebibliography}

\end{document}